\documentclass{article}
\usepackage{amssymb}
\usepackage{amsmath,bm}

\title{What is missing from Minkowski's ``Raum und Zeit'' lecture}
\author{Thibault Damour \\
\\
{\it {\small Institut des Hautes \'Etudes Scientifiques,}} \\ 
{\it {\small 35, route de Chartres, 91440 Bures-sur-Yvette}}}
\date{\small 8 July 2008}

\begin{document}

\maketitle

\begin{abstract}
This contribution tries to highlight the importance of Minkowski's \break ``Raum und Zeit'' lecture in a ``negative'' way, where {\it negative} is taken in the photographic sense of reversing lights and shades. Indeed, we focus on the ``shades'' of Minkowski's text, i.e. what is missing, or misunderstood. In particular, we focus on two issues: (i) why are Poincar\'e's pioneering contributions to four-dimensional geometry not quoted by Minkowski (while he abundantly quoted them a few months before the Cologne lecture)?, and (ii) did Minkowski fully grasp the physical (and existential) meaning of ``time'' within spacetime? We think that this ``negative'' approach (and the contrast between Poincar\'e's and Minkowski's attitudes towards physics) allows one to better grasp the boldness of the {\it revolutionary} step taken by Minkowski in his Cologne lecture.
\end{abstract}


\section{Introduction}

Minkowski's September 1908 Cologne lecture ``Raum und Zeit'' \cite{RZ} was certainly a landmark event which launched a new way of representing physical reality. Yet, some physicists, and most notably Einstein himself, reacted somewhat negatively to Minkowski's 4-dimensional reformulation of Special Relativity. Actually, Albert Einstein and Jakob Laub \cite{EL08} were among the firsts to react in print, even before the Cologne lecture, to Minkowski's first technical paper, published in April 1908 \cite{Grundgleichungen}. They expressed the concern that the 4-dimensional formalism would place ``rather great demands'' on the readers of Annalen der Physik \cite{Walter} (they also had technical objections to Minkowski's definition of a non-symmetric electromagnetic stress-energy tensor in a polarizable medium). Other physicists (notably Planck, who had been the first leading physicist to understand the conceptual novelty of Einstein's 1905 Special Relativity paper) appreciated the elegance of Minkowski's reformulation of Special Relativity, while others (and most notably Sommerfeld) came to comprehend Special Relativity in great part thanks to the 4-dimensional formalism of Minkowski. Actually, Minkowski's first technical paper \cite{Grundgleichungen} used an abstract mathematical notation\footnote{For instance, he writes Maxwell's equations as ${\rm lor} \, f = -s$ and ${\rm lor} \, F^* = 0$. Here, $s$ denotes the current, ${\rm lor}$ a contraction with the 4-dimensional gradient operator, and $*$ the (Hodge) dual ($f \sim ({\bm D}, {\bm H})$; $F \sim ({\bm E} , {\bm B})$). This notation is very close to the modern notation, $\delta f = s$, $\delta * F = 0$, in terms of the (Hodge) {\it dual}
of the Cartan exterior derivative: $\delta = * \, d \, *$. Moreover, Minkowski uses the matrix calculus to express the covariance properties of the ``matrices'' $f = (f_{\mu\nu})$ and $F = (F_{\mu\nu})$ under a change of coordinate system. E.g. he essentially writes $x = A x'$, $f' = A^{-1} f A$. [He uses $x_4 = it$ and an Euclidean metric $e = (e_{\mu\nu}) = (\delta_{\mu\nu})$: ``Einheitsmatrix''.] No wonder that such unfamiliar (and partly new) notation did not immediately appeal to physicists.} which was not of much help for physicists. Through the efforts of several physicists
(notably Sommerfeld and Max von Laue) Minkowski's abstract
formalism was translated  into a tensor calculus form 
which helped physicists to master, and use with profit, the 4-dimensional formalism\footnote{A later simplification was brought in by Einstein when he introduced his famous ``summation convention''.}. See \cite{Walter2} for a
historical discussion of the application of the 4-vector
formalism to gravitation.

\smallskip

Anyway, after a comprehensible initial reluctance caused by the unfamiliar notation of Minkowski, his formulation started to attract a lot of attention (both from mathematicians and physicists), and it is clear that the dramatic tone\footnote{``Henceforth space by itself, and time by itself, are doomed to fade away into mere shadows$\ldots$''; ``$[\ldots]$ mathematics $[\ldots]$ is able $[\ldots]$ with its senses sharpened by an unhampered outlook to far horizons, to grasp forthwith the far-reaching consequences of such a metamorphosis of our concept of nature.''; ``Thus the essence of this postulate may be clothed mathematically in a very pregnant manner in the mystic formula $3.10^5 \, {\rm km} = \sqrt{-1} \, {\rm secs}$.''}, and non-technical nature of his Cologne lecture was instrumental in capturing the imagination and interest of many scientists world-wide. In addition, the tragic and sudden death of the still young (44) Minkowski (due to an appendicitis) a few months after the Cologne lecture, added a further romantic aura (\`a la Galois) to the last publication he completed during his lifetime. Let us also note (as emphasized in \cite{Walter}) that Minkowski's Cologne lecture benefitted from a large diffusion. Within a few months it was published in three periodicals, as well as a booklet published by Teubner in 1909. In addition, French \cite{ET} and Italian translations of his lecture appeared before the end of 1909. As an example of the ``amplification effect'' that the dramatic death of Minkowski may have had on the diffusion of his new vision of space and time, let us quote the introductory note added by A.~Gutzmer, the editor of the Teubner booklet\footnote{I thank Ingrid Peeters for help in translating this introductory note.}:

\begin{quotation}
``The lecture on {\it Space and Time} delivered by Minkowski in Colo\-gne is the last creation of his genius. It was, alas, not given to him to accomplish his bold project: to set up a Mechanics where time unites itself with the three dimensions of space. On January 12, 1909, a tragic fate snatched, in the prime of life, an author equally esteemed for his human and scientific qualities, away from the affection of his family and his friends.

The keen and well-deserved interest raised by his lecture had greatly pleased Minkowski. He had wished to make his reflections accessible to a larger audience by means of a special publication. It is a painful duty of piety and friendship for the Teubner editions, and the undersigned, to fullfil, by the present publication, this last wish of the departed.

\hfill A. Gutzmer.''
\end{quotation}

But evidently, beyond the de-amplification and amplification effects caused by the various factors recalled above (a very abstract notation in the April 1908 technical paper, the somewhat theatrical tone of Cologne's non-technical expos\'e, the romantic aura added by a sudden and untimely death), there is no doubt that it is the conceptual novelty of Minkowski's vision which stirred a well-deserved attention, and launched a new way of thinking about physical reality. We all know how fruitful Minkowski's Space-Time concept has been. The most brilliant successes (and puzzles!) of modern physics are rooted in the various avatars of his concept: from General Relativity to Relativistic Quantum Field Theory, and the difficulty of uniting them (that String Theory aims at solving). I assume that other contributions to this commemorative issue of Annalen der Physik will cover the
legacy of Minkowski's vision. I wish here to approach Minkowski's Cologne lecture from a particular angle.

\smallskip

Namely, I will focus on several very important aspects of what we now associate with Minkowski's Space-Time which are {\it missing} in the Cologne lecture. I hope that this {\it negative} approach to Minkowski's lecture (``negative'' being taken in the photographic sense) might be useful for highlighting, by contrast, its (positive) content, and for appreciating some of the subtleties underlying the concept of ``scientific revolution''.

\section{Why are Poincar\'e's contributions not quoted by Minkowski?}

The most blatant omission\footnote{This omission has
been noted and discussed in \cite{Walter}.} in Minkowski's Cologne lecture is the name of Poinca\-r\'e. Indeed, while Minkowski mentions W.~Voigt, [A.A.]~Michelson, H.A.~Lorentz, A.~Einstein, M.~Planck, I.R.~Sch\"utz, A.~Li\'enard, E.~Wiechert, and K.~Schwarz\-schild\footnote{Let us recall, in this respect, that W.~Voigt, E.~Wiechert and K.~Schwarzschild were colleagues of Minkowski in G\"ottingen. For information concerning the scientific life in G\"ottingen, we relied on the well-documented book \cite{Leveugle} (without, however, supporting the strange complot theory advocated by the author).}, as well as his own technical paper \cite{Grundgleichungen}, he never mentions Poincar\'e. Yet, several of the key mathematical results in the 4-dimensional formulation of Special Relativity had been obtained by Poincar\'e in 1905 \cite{P05,P06}, notably:
\begin{itemize}
\item[(i)] the fact that Lorentz transformations leave invariant the quadratic form $x^2 + y^2 + z^2 - c^2 t^2$ (and the fact that they form a group), 
\item[(ii)] the fact that the four electromagnetic potentials\footnote{For conciseness and readability,
we do not respect here the notation of Poincar\'e.
Poincar\'e writes the 4-dimensional quantities in components, $x^{\mu} = (x,y,z,t)$; $A^{\mu} = (F,G,H,\psi)$; $J^{\mu} = (\rho \, \xi , \rho \, \eta , \rho \, \zeta , \rho)$; $F^{\mu\nu} = (\alpha , \beta , \gamma ; f,g,h)$. He uses, however, a streamlined notation which makes clear which quantities transform in the same way, and which quantities are invariant.} $A^{\mu}$ (in Lorenz gauge, $\partial_{\mu} \, A^{\mu} = 0$) transform (under the Lorentz group) in the same way as $x^{\mu}$ and the `4-current' $J^{\mu}$,
\item[(iii)] the fact that $F_{\mu\nu} \, F^{\mu\nu}$ and $\varepsilon^{\mu\nu\rho\sigma} F_{\mu\nu} \, F_{\rho\sigma}$ are Lorentz-invariant, 
\item[(iv)] the necessity of endowing each electron (modelled as a sphere) with an internal tension (or negative pressure) to be compatible with the ``Postulat de Relativit\'e'', i.e. the impossibility of experimentally detecting the absolute motion,
\item[(v)] the resulting Lorentz-invariant form of the Lagrangian for the electron dynamics, $L \propto \sqrt{1 - {\bm v}^2/c^2}$, 
\item[(vi)] the introduction of a `Wick-rotated' time coordinate $x^4 = ict$ (denoted $ t \sqrt{-1}$ by Poincar\'e)
and the associated technical use of a `4-dimensional space' (an ``espace \`a 4 dimensions'') in which Lorentz transformations become Euclidean rotations around the origin of $x^{\mu}$ ($\mu = 1,2,3,4$), and, last but not least,
\item[(vii)] the construction of Lorentz-covariant generalizations of Newton's $1/r^2$ gravitational force.
\end{itemize}

\noindent In the latter construction, Poincar\'e makes an essential use of 4-dimensional geometry, both mathematically (construction of invariants and 4-vectors associated with two separate spacetime points $x_0^{\mu}$, $x_1^{\mu}$ and their infinitesimal displacements $dx_0^{\mu} = u_0^{\mu} d\tau_0$, $dx_1^{\mu} = u_1^{\mu} d\tau_1$), and physically (consideration, among other Lorentz-invariant possibilities, of an action-at-a-distance propagating\footnote{Let us mention in this respect that Poincar\'e introduced the concept of gravitational wave (``onde gravifique'') already in his brief June 1905 summary \cite{P05}.} between $x_1^{\mu}$ and $x_0^{\mu}$ along the retarded light-cone $\eta_{\mu\nu} (x_1^{\mu} - x_0^{\mu}) (x_1^{\nu} - x_0^{\nu}) = 0$). Note that we are here rephrasing the achievements of Poincar\'e in modern (post-Minkowski) notation. The original text of Poincar\'e is somewhat less transparent. In particular, though Poincar\'e explicitly introduces the infinitesimal displacements $dx_0^{\mu} (= u_0^{\mu} d\tau_0)$, $dx_1^{\mu} (= u_1^{\mu} d\tau_1)$ (linked to the 3-velocities $v_0^i = dx_0^i / dt_0$, $v_1^i = dx_1^i / dt_1$), he introduces only {\it implicitly} the proper times $d\tau_a = \sqrt{-\eta_{\mu\nu} \, dx_a^{\mu} \, dx_a^{\nu}} = \sqrt{1 - {\bm v}_a^2} \ dt_a$ ($a=0,1$; using, like Poincar\'e, $c=1$ for simplicity), and the corresponding 4-velocities $u_0^{\mu}$, $u_1^{\mu}$, through several normalization factors. For instance, he writes the 4 basic invariant scalar products
\begin{equation}
\label{eq2.1}
\eta_{\mu\nu} (x_1^{\mu} - x_0^{\mu}) (x_1^{\nu} - x_0^{\nu}) \, , \ - \eta_{\mu\nu} (x_1^{\mu} - x_0^{\mu}) \, u_0^{\nu} \, , \ -\eta_{\mu\nu} (x_1^{\mu} - x_0^{\mu}) \, u_1^{\nu} \, , \ -\eta_{\mu\nu} \, u_0^{\mu} \, u_1^{\nu} \, ,
\end{equation}
in the forms
\begin{equation}
\label{eq2.2}
\sum x^2 - t^2 \, , \ \frac{t-\sum x \, \xi}{\sqrt{1-\sum \xi^2}} \, , \ \frac{t-\sum x \, \xi_1}{\sqrt{1-\sum \xi_1^2}} \, , \ \frac{1-\sum \xi \xi_1}{\sqrt{(1-\sum \xi^2) \
(1-\sum \xi_1^2)}} \, ,
\end{equation}
where Poincar\'e's $(x,y,z,t)$ denote what we denoted $x_1^{\mu} - x_0^{\mu}$, while $\xi$ denotes the first component of the 3-velocity $v_0^i = dx_0^i / dt_0$ ($i=1,2,3$), and $\xi_1$ the first component of $v_1^i = dx_1^i / dt_1$.

\smallskip

The omission of the name of Poincar\'e cannot be explained by Minkowski's unawareness of Poincar\'e's  achievements. Indeed, Minkowski quoted (in a positive and detailed manner) the works of Poincar\'e on two occasions in 1907-1908. First, on November, 5, 1907 (i.e. nearly one year before the lecture in Cologne) Minkowski gave a lecture to the G\"ottinger Mathematischen Gesellschaft. The written text of this lecture was published posthumously in 1915 (with the title:
 ``Das Relativit\"atsprinzip'')  , through the efforts of Sommerfeld \cite{M15}. This is the first account of Minkowski's ideas on Special Relativity and 4-dimensional geometry. It is striking that, in this text, Poincar\'e's name is among the most cited ones: more precisely, the three most cited names are Planck (cited eleven times), Lorentz (cited ten times) and Poincar\'e (cited six times). By contrast, Einstein's name appears only twice! It is also interesting to note that, in this text, Minkowski credits Poincar\'e (together with Einstein and Planck) for having elaborated the ``Postulat der Relativit\"at'' in a form ``understandable by mathematicians''. He also credits Poincar\'e for having (after Lorentz) discovered the invariance of the equations of electrodynamics under a {\it group}. At the end, he explicitly mentions how Poincar\'e generalized Newton's gravitational force to a relativistic form by using several possible invariants of the Lorentz group. However, one also finds a way of mentioning Poincar\'e which anticipates the future downplaying of Poincar\'e contributions by Minkowski. Namely, Minkowski writes:

\begin{quotation}
``Besides the fact that they are independent of the choice of rectangular coordinate system in space, these
fundamental [electrodynamic] equations possess another symmetry, which has not yet been made manifest by the usual notations.  I will here expose this symmetry
from the start (which none of the
cited authors did, not even Poincar\'e), by using a form of the equations which makes it absolutely transparent [durchsichtig].''
\end{quotation}

After which, Minkowski introduces the notation $x_1 = x$, $x_2 = y$, $x_3 = z$ and $x_4 = it$ (in units where $c=1$) for the coordinates in space and time, and refers to a ``four-dimensional manifold'' (``vierdimensionalen Mannigfaltigkeit''). He then represents the ``electrodynamic state in space at any time'' (``elektromagnetische Zustand im Raum zu jeder Zeit'') by a ``vierdimensionalen Vektor'' $\psi_1 , \psi_2 , \psi_3 , \psi_4 = i \, \Phi$ (which is the electromagnetic 4-potential $A_{\mu}$ in the `Pauli metric' $\delta_{\mu\nu}$). He then writes explicitly the Lorentz electrodynamic equations (in Lorenz gauge), for instance (with, in Minkowski's notation $j = 1,2,3,4$ and $\rho_j = 4$-current)
\begin{equation}
\frac{\partial \, \psi_1}{\partial \, x_1} + \frac{\partial \, \psi_2}{\partial \, x_2} + \frac{\partial \, \psi_3}{\partial \, x_3} + \frac{\partial \, \psi_4}{\partial \, x_4} = 0 \, ,
\end{equation}

\begin{equation}
\Box \, \psi_j = - \rho_j \, ,
\end{equation}

\begin{equation}
\psi_{jk} = \frac{\partial \, \psi_k}{\partial \, x_j} - \frac{\partial \, \psi_j}{\partial \, x_k} = - \psi_{kj} \, .
\end{equation}

He does not use a tensor-calculus notation, and writes explicitly the sums over repeated indices when he needs them (as exemplified by the Lorenz-gauge condition above). Probably, Poincar\'e, had he been aware of Minkowski's November 1907 lecture, would not have considered that it contained any significantly new results. Indeed, Poincar\'e was used to denoting a 3-dimensional vector as an explicit triplet of successive  letters, such as $(f,g,h)$; $(\alpha , \beta , \gamma)$; $(u,v,w)$; $(\xi , \eta , \zeta)$. Therefore, Poincar\'e's notation (see above), say, $(\rho \, \xi , \rho \, \eta , \rho \, \zeta, \rho)$ for the (convection) electric 4-current $(J^{\mu})$ was as clear to him as the 4-index notation $\rho_j$ used by Minkowski.
By contrast, it seems that the fact that the use of spacetime indices $j,k = 1,2,3,4$ led to making more ``transparent'' the 4-dimensional symmetry was instrumental in psychologically convincing Minkowski that he was breaking new ground, beyond Poincar\'e (see Minkowski's citation above). Let us note in this respect that while in France it was usual at the time to denote vectors as triplets of letters, $(x,y,z)$; $(u,v,w)$; etc, Germany and England used, at once, two different notations: the explicit one, say $(X,Y,Z)$, together with the abstract vector one, say ${\mathfrak E}$ (to use a notation employed in Maxwell's treatise on electricity and magnetism)\footnote{This notational difference might in turn have been rooted in mathematical advances in abstract algebra which took place mainly in England and Germany, such as the developement of quaternions (used in England, starting with Maxwell, to write the electrodynamic equations in compact form), and the one of Grassmann algebras. Note, however, that Einstein, in his first papers, denote the electric and magnetic fields as explicit triplets: $(X,Y,Z)$; $(L,M,N)$. By contrast, H.A.~Lorentz uses, when he can, a more condensed notation, e.g. ${\rm div} \, {\bm D} = \rho$, ${\rm div} \, {\bm H} = 0$, and denotes their components as $(D_x , D_y , D_z)$, $(H_x , H_y , H_z)$, when he needs to be explicit.}. In addition to the psychologically convincing (for Minkowski) ``transparency'' brought by the use of an explicit 4-dimensional index notation, his November 1907 lecture contains one useful technical advance, namely the concept now called a 2-form (such as $F = \frac{1}{2} \, F_{\mu\nu} \, dx^{\mu} \wedge dx^{\nu}$), which he calls a ``Traktor'', i.e. a six-component spacetime object $(p) = (p_{jk})$ (with $p_{jk} = - p_{kj}$) such that $p_{jk} \, x^j \, y^k$, or, as he writes explicitly,
\begin{equation}
p_{23} (x_2 \, y_3 - x_3 \, y_2) + p_{31} (x_3 \, y_1 - x_1 \, y_3) + \cdots + p_{34} (x_3 \, y_4 - x_4 \, y_3) \, ,
\end{equation}
is invariant under 4-dimensional rotations (when $x_j$ and $y_j$ transform as 4-vectors). In addition, he emphasizes that both the six components of the electromagnetic field $\psi_{jk}$, and those of the polarization tensor (entering the electrodynamics within electrically and magnetically polarizable media), are spacetime `traktors'.

Nonetheless, I find it probable that Poincar\'e (who
did not care about notations) might have been unimpressed
by Minkowski's discussion of the specific ``tensor variance''
of various quantities, such as $\psi_{jk}$. In this respect,
let us stress that one must distinguish between the (useful)
technology of ``tensor calculus'' (using, say, the index notation), and the general mathematical understanding of
tensors. The (later) example of \'Elie Cartan suggests that
French mathematicians had (in keeping with their different
habit about the way of denoting vectors, see above)
a general way of thinking about tensors which did not need
to rely on any ``index notation''. For instance, \'E. Cartan
defines a general tensor (under any group) as
a ``linear representation'' viewed in concrete terms, i.e.
a collection of
numbers $(u_1,u_2, \ldots, u_r)$ which transform linearly
and homogeneously under each group operation, in a way
consistent with the group law (see, e.g., chapter II in \cite{Cartan}).
 From this point of view, the work
of Poincar\'e (together with the previous work of Lorentz)
had already shown that the electromagnetic potentials
$ (u_1,u_2,u_3,u_4) = (F,G,H,\psi) (=A^{\mu}) $, the 4-current $ (u'_1,u'_2,u'_3,u'_4) =(\rho \, \xi , \rho \, \eta , \rho \, \zeta , \rho) (= J^{\mu})$,
and the electromagnetic field $(u''_1,u''_2,u''_3,u''_4,u''_5,u''_6)=
(\alpha , \beta , \gamma ; f,g,h) (= F^{\mu\nu})$ were tensors
of $SO(3,1)$. The important point to note is that this
general definition needs neither to rely on any specific
index notation, nor to presuppose any tensor calculus.
On the other hand, it is true that, in the specific case of
tensors under (say) a rotation group, the index notation, together with the associated index calculus, is a useful tool
for combining some given tensors into new tensors
(`tensor calculus' or `tensor algebra'). This lack
of a good associated calculus explains why the mathematical
discovery of ``spinors'' by \'E. Cartan
as early as 1913 \cite{Cartan1913} (as certain fundamental representations of the rotation group) dropped into oblivion
until the work of physicists (notably Pauli and Dirac), interested in the quantum mechanics of particles with spin,
rediscovered them in the late 1920's, together with an
associated ``spinor calculus'' (index notation, matrix calculus, Clifford algebra, etc.).

\smallskip

The next step in Minkowski's deepening of 4-dimensional geometrical concepts\footnote{Here, we focus on Minkowski's conceptual achievements concerning spacetime geometry. In addition to these, Minkowski clarified the relativistic formulation of the electrodynamics of polarizable media  (involving two different 2-forms, say $F_{\mu\nu} = \partial_{\mu} A_{\nu} - \partial_{\nu} A_{\mu}$ and $f^{\mu\nu}$, such that $\partial_{\nu} f^{\mu\nu} = -J^{\mu}$), and introduced the concept of (spacetime-covariant) electromagnetic stress-energy tensor $t_{\mu}^{\nu} = F_{\mu\sigma} f^{\nu\sigma} - \frac{1}{4} \, \delta_{\mu}^{\nu} F_{\rho\sigma} f^{\rho\sigma}$. However, the fact that this tensor was {\it non-symmetric} within polarizable bodies initiated a long series of arguments (starting with the work of Einstein and Laub \cite{EL08} mentioned above).} probably happened between November 5, 1907 and December 21, 1907, when he presented to the G\"ottingen Royal Society of Sciences a technical article entitled ``The fundamental equations of electromagnetic processes in moving bodies'' \cite{Grundgleichungen}. This work contains (among other important advances) one key new insight: the concept of {\it spacetime line} (``Raum-Zeitlinie'', which will be called by him {\it worldline}, ``Weltlinie'', in the Cologne lecture) together with the concept of {\it proper time} (``Eigenzeit'') along it. These concepts are not present in Minkowski's November 1907 lecture. Actually, as emphasized in Ref.~\cite{Walter2}, the latter lecture \cite{M15} contains incorrect considerations concerning the definition of the `4-velocity' of a mass point. Indeed, in \cite{M15} Minkowski denotes the ordinary 3-velocity $d{\bm x} / dt$ as ${\bm w} \equiv d{\bm x} / dt$, states that $\vert {\bm w} \vert$ must stay smaller than unity (using $c=1$), and claims that
\begin{equation}
w_1 = w_x \, , \ w_2 = w_y \, , \ w_3 = w_z \, , \ w_4 = i \, \sqrt{1-{\bm w}^2}
\end{equation}
define a suitable 4-dimensional velocity vector satisfying the Lorentz-invariant constraint
\begin{equation}
w_1^2 + w_2^2 + w_3^2 + w_4^2 = -1 \, .
\end{equation}
The latter claim is, however, technically incorrect. Note, in particular, that $w_1^2 + w_2^2 + w_3^2 + (i \, \sqrt{1-{\bm w}^2})^2 = {\bm w}^2 - (1-{\bm w}^2) = -1 + 2 {\bm w}^2$ is not equal to $-1$. It seems probable to me that Minkowski understood his error and corrected it by having a closer look at Poincar\'e's 1906 paper (which, as said above, implicitly contained the correct answer). Indeed, Minkowski ends his November lecture by a few remarks on the ``big question'' (``grosse Frage'') of putting gravitation in line with the Relativit\"atsprinzip. This final paragraph contains no equations, but refers abundantly to the work of Poincar\'e \cite{P06}, and ends by the promise of giving a detailed report on this issue on another occasion. As all scientists know, preparing and giving a lecture often forces you to come back and think about the topic of the lecture. It is therefore quite likely that, after his November lecture, Minkowski went back to the final section (\S9) of the 1906 Poincar\'e paper and put more effort in understanding what Poincar\'e had done, and in translating it into a more transparent spacetime language.

\smallskip

At this point, I should indeed point out to the readers who have never tried to read the 1906 Rendiconti paper of Poincar\'e that this article is rather long (47 pages) and technically quite complex, especially in the final section devoted to some ``hypoth\`eses sur la Gravitation''. Indeed, many of the key new results of Poincar\'e on Relativity are contained in this final section, but in a rather untransparent and unpedagogical form. For instance, the crucial fact that Lorentz transformations preserve the quadratic form $x^2 + y^2 + z^2 - t^2$ is asserted in passing as if it were well known (``An arbitrary transformation of this group can always be decomposed into [a scaling], and into a linear transformation leaving unchanged the quadratic form $x^2+y^2+z^2-t^2$";
and later:``We know that$\ldots$'', ``Nous savons que$\ldots$''), and the novel introduction of a 4-dimensional `Wick-rotated' $(t \, \sqrt{-1})$ Euclidean space is also done in passing, with the brief comment: ``We see that the Lorentz transformation is just a rotation of this space around the origin, regarded as being fixed.'' Essentially, the introduction of this 4-dimensional geometrical representation is used by Poincar\'e only as a technical tool to construct Lorentz scalars and Lorentz vectors from the geometrical configuration defined by the attracted point $(x_0^{\mu})$ and the attracting one $(x_1^{\mu})$, together with their infinitesimal spacetime displacements $(dx_0^{\mu} , dx_1^{\mu})$, connected to their respective velocities $(dx_0^i / dx_0^0 = v_0^i , dx_1^i / dx_1^0 = v_1^i)$\footnote{I am using here post-Minkowski notations instead of those used by Poincar\'e.}. In reading more closely this section\footnote{At this stage, I would like to emphasize that, contrary to what many historians of science seem to assume when discussing the information contained in scientific papers, many (if not most?) scientists practically never read in detail, from the beginning to the end, any scientific paper, even if it is central to their own interest and research. More often than not, they just glance through the papers written by others, trying to capture fast its (generally few) new conceptual points and technical results, and ``translating'' them into their usual way of thinking or of calculating.}, Minkowski may have then realized that the construction of Poincar\'e was illuminated by extending his geometrical configuration $(x_0^{\mu} , dx_0^{\mu} ; x_1^{\mu} , dx_1^{\mu})$ into two full {\it spacetime lines} $x_0^{\mu} (\tau_0)$, $x_1^{\mu} (\tau_1)$, whose {\it tangents} at $x_0^{\mu}$ and $x_1^{\mu}$ would be proportional to the infinitesimal displacements $dx_0^{\mu} , dx_1^{\mu}$ considered by Poincar\'e. This also illuminated the ``good'' geometrical way of normalizing the tangent vectors: namely, to use as parameters $\tau_0 , \tau_1$ the invariant spacetime length, $d\tau_a^2 = -\eta_{\mu\nu} \, dx_a^{\mu} dx_a^{\nu}$, for which Minkowski introduces the name of {\it proper time} (``Eigenzeit''). Let us emphasize again that Poincar\'e had already correctly normalized his invariants (\ref{eq2.2}) by the factors $(1-{\bm v}_a^2)^{-1/2}$ appropriate to the
proper-time 4-velocity $u_a^{\mu}$ (see the corresponding Eqs. (\ref{eq2.1})), without, however, explaining in detail what he was doing\footnote{As often, Poincar\'e found such technical details so easy to perform that he did not bother to explain them at length. We note, however, that the sentence he writes before deriving his invariants (\ref{eq2.2}) (Eq. (5) in his text \cite{P06}) is a mathematically complete (though rather cryptic) way of saying he uses $u_a^{\mu} = dx_a^{\mu} / d\tau_a$. Indeed, he says that ``one must $\ldots$ look for [those invariant combinations of $x_1^{\mu} - x_0^{\mu} , dx_0^{\mu} , dx_1^{\mu}$] which are homogeneous of degree zero both with respect to $\delta x , \delta y , \delta z , \delta t$ [i.e. $dx_0^{\mu}$] and with respect to $\delta_1 x , \delta_1 y , \delta_1 z , \delta_1 t$ [i.e. $dx_1^{\mu}$].'' This is an implicit way of saying that he uses the ratios $\delta x / \sqrt{\delta t^2 - \delta x^2 - \delta y^2 - \delta z^2}$, etc., i.e. $dx_a^{\mu} / d\tau_a$.}.

\smallskip

If this conjectural reconstruction of the way Minkowski realized, by reading closely Poincar\'e, that the concepts of spacetime line (i.e. worldline in the Cologne lecture) and proper time were crucial in geometrically representing the dynamics of interacting point particles is correct, it is all the more surprising that Minkowski started downplaying the contributions of Poincar\'e in his December 1907 paper \cite{Grundgleichungen}. While, as we said above, Minkowski frequently, and rather warmly, quotes Poincar\'e in his November 1907 lecture \cite{M15}, he quotes him only twice in his December 1907 paper, and in a rather derogatory manner. Indeed, at the beginning, he mentions Poincar\'e only for having given the name of ``Lorentz transformations'' to the covariance properties of ``the theory of Lorentz'', without ever mentioning the new results of Poincar\'e (first proof of the covariance of Lorentz' theory in presence of electric currents $J^{\mu}$, first proof of the group character of the Lorentz transformations). Then, he mentions results brought by Poincar\'e without citing him (introduction of Euclidean time $it$, proof of a ``theorem of relativity'', invariance of $F_{\mu\nu} \, F^{\mu\nu}$ and $\varepsilon^{\mu\nu\rho\sigma} F_{\mu\nu} \, F_{\rho\sigma}$, etc). Finally, at the end, when discussing his own (much less general than Poincar\'e's) way of reconciling gravitation with the ``postulate of relativity'', he cites Poincar\'e contributions only in a rather strange footnote: ``In a way completely different from the one I employ here, H. Poincar\'e \cite{P06} has tried to adapt the law of Newtonian attraction to the postulate of relativity.''

\smallskip

It seems that, somehow, Minkowski's full realization of the elegant 4-dimen\-sional geometrical formulation of Special Relativity pioneered by Poincar\'e in a more technical way, and using less transparent notations, put him in the psychological mood of downplaying, and ultimately completely neglecting (in the Cologne lecture) Poincar\'e's contributions. This psychological reaction had probably several different roots.

\smallskip

Let us first recall the general context of the relations between Germany and France, and between German mathematicians and French ones. During the Franco-Prussian war (July 1870-May 1871), Germany had invaded France, and had taken possession of two regions of France: Alsace and a large part of Lorraine (the latter being the native region of Poincar\'e). Felix Klein (born in 1849) was visiting (at the same time as Sophus Lie) the Paris mathematical school in 1870, where he discovered and studied group theory in the newly published treatise of Camille Jordan (``Trait\'e des substitutions''). This treatise, as well as his interaction with Lie, was crucial to the line of research later pursued by Klein (notably his 1872 ``Erlangen programme''). The Franco-Prussian war interrupted Klein's stay in Paris. Klein (who was prussian and extremely patriotic) ``rushed home to volunteer for the army'' (\cite{Reid} p.137) during the war against France. In the early 1880's, Klein engaged in a sharp (but friendly) competition with the young Poincar\'e (born in 1854) about automorphic functions. The final result of this competition was essentially a draw, with a slight advantage for Poincar\'e. However, the mental pressure from the competition caused a serious nervous breakdown of Klein. I am recalling this as background material because Hilbert and Minkowski, the new stars of the G\"ottingen mathematical group in the early 1900's, were both prussians, and had been attracted to G\"ottingen by Klein. The latter had previously (circa 1885) advised Hilbert to go to Paris because it would be ``most stimulating and profitable'' for him, especially if he ``could manage to get on the good side of Poincar\'e'' \cite{Reid}. When Hilbert visits Paris, his old friend Minkowski writes to him, commenting that he is ``in enemy territory''. Let us also mention the comment of Hurwitz (a common mentor and friend to Hilbert and Minkowski since their youth in K\"onigsberg, \cite{Reid} p.14):

\begin{quotation}
``I fear the young talents of the French are more intensive than ours, so we must master all their results to go beyond them.'' (\cite{Reid}, page 21.)
\end{quotation}

The latter citation of Hurwitz\footnote{Let us also recall that Hurwitz was professor at the Z\"urich Polytechnikum, where he had attracted Minkowski (in 1896), before Klein could secure a position for him in G\"ottingen. Both Hurwitz and Minkowski taught at the Polytechnikum when Einstein studied there. Though the above remark of Hurwitz was addressed to Hilbert (before his stay in Paris), it is most likely that he made similar remarks to Minkowski (e.g. during their common time in Z\"urich).} is the closest approximation I could find to an explanation of Minkowski's downgrading of Poincar\'e's work. When Minkow\-ski realized he could go beyond Poincar\'e, both technically and conceptually\footnote{And {\it diagrammatically}. Indeed, one of the key new results of the Cologne lecture consisted in the first introduction of (2-dimensional) spacetime diagrams. See \cite{Galison} for a discussion of the importance of visual thinking in Minkowski's work.}, after mastering his rather opaque technical methods, his hubris pushed him to downplay Poincar\'e's contributions.

This downplaying may have been facilitated by the following
conjectural reconstruction of the way Minkowski discovered
the fact that Lorentz transformations are ``rotations'' in
space-time. Max Born, who had attended the seminar on electron
theory co-organized by Minkowski and Hilbert in the summer
semester of 1905, wrote\footnote{I thank Scott Walter for
giving me the original German citation of Born, and
Friedrich Hehl for advice about its translation into English.},
many years later (in 1959) \cite{Born1959}:

\begin{quotation}
`` I remember  that Minkowski occasionally
alluded to the fact that he was
engaged with the Lorentz transformations,
and that he was on the track of new interrelationships.
[Ich erinnere mich, da{\ss} Minkowski gelegentlich Andeutungen
machte, da{\ss} er sich mit den Lorentz-Transformationen besch{\"a}ftigte und
neuen Zusammenh{\"a}ngen auf der Spur sei.]''
\end{quotation}

My conjecture is that Minkowski, helped by his background
reading of some of the works of Lorentz and Poincar\'e
(which, however, {\it did not include} their most recent
contributions of 1904-1905; see \cite{Leveugle} and references
therein) had discovered by himself, in the summer of 1905
(without knowing about the 1905 papers of Poincar\'e) the fact
that Lorentz transformations preserve the quadratic form
$ - c^2 t^2 + x^2 (+ y^2 + z^2)$. If that reconstruction is
correct, he must have been all the more eager, when he later
realized that he had been preceded by Poincar\'e, to find
reasons for downplaying Poincar\'e's work.

 Years later, some  fairer scientists tried to correct this situation. In particular, Sommerfeld added some Notes to a republication of Minkowski's Raum und Zeit lecture in the well-known booklet ``The Principle of Relativity'' \cite{RZ} in which he acknowledges (though only partially) Poincar\'e's contributions. A better job was done by the young Pauli (apparently under the insistence of Felix Klein
 himself \cite{Enz}) in his famous (book-size) article on the theory of relativity for Klein's Mathematical Encyclopedia \cite{Pauli21}.

\section{Did Minkowski really think of time as a mere shadow?}
\setcounter{equation}{0}

The most quoted sentences from Minkowski's Cologne lecture are those concerning the subsuming of space and time under a new, four-dimensional reality, notably:

\begin{quotation}
``Henceforth, space by itself, and time by itself, are doomed to fade away into mere shadows, and only a kind of union of the two will preserve an independent reality.'' [...]

``Three-dimensional geometry becomes a chapter in four-dimen\-sional physics. Now you know why I said at the outset that space and time are to fade away into shadows, and only a world in itself will subsist.''
\end{quotation}

Far from me to try to downplay the conceptual revolution initiated by Minkowski. However, if one reads the entire text of the Cologne lecture, one does not go away with the feeling that Minkowski took the new spacetime concept as being {\it existentially} relevant to us, as human beings. Though Minkowski certainly went much farther than Poincar\'e in taking seriously the 4-dimensional geometry as a new basis for a physico-mathematical representation of reality, it does not seem that he went, philosophically and existentially, as far as really considering `the flow of time' as an illusory shadow. By contrast, let us recall that the old Einstein apparently did take seriously, at the existential level, the idea that `time' was an illusory shadow, and that the essence of (experienced) reality was timeless. For instance, in some letters to his old (Polytechnikum) friend Michele Besso, he writes \cite{EinsteinBesso}

\begin{quotation}
`` $\ldots$ you do not take seriously the four-dimensionality of Relativity, but you consider the present as the only reality.'' (letter 185; 13 July 1952)

``You cannot get used to the idea that subjective time, with his ``now'', has no objective meaning. See Bergson!'' (letter 197; 29 July 1953)
\end{quotation}

An even clearer assertion of this idea is contained in the famous letter of condolences written by Einstein after Besso's death:

\begin{quotation}
``Now he has departed from this strange world a little ahead of me. That signifies nothing. For us, physicists in the soul, the distinction between past, present and future is only a stubbornly persistent illusion.'' (letter 215; 21 March 1955)
\end{quotation}

It is an interesting question to understand how, when and through the minds of whom, the physico-mathematical concept of Minkowski spacetime, with its ``shadowy'' times and spaces, came to be existentially experienced in this way. I am not sure of the correct answer, but I would like to offer a few thoughts.

\smallskip

First, I would like (after many others, see \cite{Walter} and references therein) to stress that Minkowski probably did not really comprehend the conceptual novelty of Einstein's June 1905 paper on Special Relativity, and especially the results therein concerning {\it time}. Indeed, in his Cologne lecture Minkowski says that, while Einstein ``deposed [time] from its high seat'', ``neither Einstein nor Lorentz made any attack on the concept of space'', by which he meant that Einstein and Lorentz did not realize (as Minkowski geometrically shows in his two-dimensional spacetime diagrams) that the spatial slice $x'$ (i.e. $t' = 0$) associated with a (relatively) `moving' observer differed from the spatial slice $x$ (i.e. $t=0$) associated to the originally considered observer. However, this was precisely one of the key new insights of Einstein, namely the {\it relativity of simultaneity}! In addition, when Minkowski introduces the (geometrically motivated) concept of proper time, he does not seem to fully grasp its physical meaning. However, this is the second key new insight brought in by Einstein concerning time, namely the fact (explicitly discussed by Einstein) that, when comparing a moving clock to one remaining at rest (and marking the corresponding `rest' coordinate time $t$), the moving clock will mark (upon being reconvened with the sedentary clock) the time\footnote{Einstein only writes (after expanding it to
first order in $v^2/c^2$)
$\tau = t \, \sqrt{1-{\bm v}^2 / c^2}$ because he
explicitly discusses only clocks moving with a constant $\vert {\bm v} \vert$, though with arbitrarily varying direction, so as to allow for a spatially closed loop, possibly along a ``continuously curved line''.}
\begin{equation}
\tau = \int dt \, \sqrt{1-{\bm v}^2 / c^2} \, ,
\end{equation}
i.e. Minkowski's proper time. It seems that Minkowski was not aware of this. This is another example of a scientist misreading a paper which he knew, however, to be central to his research topic! In that case (contrary to the case of Poincar\'e's paper), Minkowski had the excuse of being a mathematician reading a physics paper.

\smallskip

After these preliminary remarks aimed at showing that Minkowski did not fully grasp the physical meaning of what he was doing, let us come back to the question of who first ``thought'' the `block time' of the spacetime picture as implying a `timeless' physical, and existential, `reality'\footnote{Note that we are limiting our considerations to the `descendants' of Minkowski's work. Another interesting question would be to consider its `ancestry'. In particular, whether Minkowski was helped by some (direct or indirect) knowledge of some literary works, such as H.G.~Wells' 1895 ``Time Machine'' [whose first pages are a rather clear description of (pre-einsteinian) spacetime] or C.H.~Hinton's 1880 essay ``What is the fourth dimension?'' (cited in \cite{Davies} as a predecessor of the concept of `block time').}. One of the first accounts I found, which goes somewhat beyond Minkowski, is contained in Einstein's 1916 popular book on Relativity \cite{E16}. In the Appendix II, entitled, ``Minkowski's Four-dimensional Space (``World'') [Supplementary to Section XVII]'', Einstein characterizes the change from the 3-dimensional-space-plus-time picture to Minkowski's ``world'' as follows:

\begin{quotation}
``From a ``happening'' in three-dimensional space, physics becomes, as it were, an ``existence'' in the four-dimensional ``world''.''
\end{quotation}

After this, I did not find evidence for the ``sinking'' of Minkowski's picture into the consciousness of scientists until the late 1940's and early 1950's. At that time, one might mention G\"odel's thoughts about the (Kantian) ideality of time (and its confirmation from the existence of closed time-like curves in the 1949 ``G\"odel universe''), Weyl's\footnote{Hermann Weyl had a philosophical bend, and his famous book on Relativity contains detailed conceptual discussions about space and time. However, I did not find there clear statements concerning the concept of `block time'. The closest citation I found (which is, however, distinct from the idea of `block time') reads: ``This world is a four-dimensional continuum which is neither ``space'', nor ``time''; it is only the consciousness which, by moving in a region of this world, registers as ``history'' a section which comes towards it, and leaves it behind, i.e. as a process which unfolds in space and develops in time.'' (end of chapter III of \cite{Weyl21}).} 1949 book ``Philosophy of Mathematics'' which (as quoted in \cite{Davies}) contains sentences such as

\begin{quotation}
``The world does not happen, it simply is.''
\end{quotation}

\noindent and various texts of Einstein of the early 1950's. Besides the letters to Besso quoted above, one might also mention\footnote{I am grateful to John Stachel for bringing this reference to my attention.} the fifth appendix that Einstein added (in June 1952) to the fifteenth (English) edition of his popular book on Relativity, entitled ``Relativity and the Problem of Space''. There, Einstein comments Minkowski's four-dimensional structure as follows:

\begin{quotation}
``Since there exist in this four-dimensional structure no longer any sections which represent ``now'' objectively, the concepts of happening and becoming are indeed not completely suspended, but yet complicated. It appears therefore more natural to think of physical reality as a four-dimensional existence, instead of, as hitherto, the {\it evolution} of a three-dimensional existence.''
\end{quotation}

To end this rather incomplete discussion, let us mention that it would be interesting to extend the study of the legacy
of the concept of relativistic spacetime to artists. Among them, Marcel Proust's ideas on Time, for instance, have indeed been somewhat influenced by einsteinian (and minkowskian) ideas (see the discussion in chapter 2 of \cite{D06}). There are probably many other examples.

\section{Other remarkable omissions}
\setcounter{equation}{0}

Among the other omissions of Minkowski, I find two worth mentioning.

\subsection{No mention of Klein's Erlangen programme}

One of the central focus of Minkowski's Cologne lecture is the {\it group} structure underlying Special Relativity (denoted as $G_c$ by Minkowski), its relation to the group structure of Newtonian mechanics ($G_{\infty} = \lim_{c \to \infty} G_c$), and the fact that mathematicians are the best armed for detecting and exploiting such group structures. I therefore find rather surprising that Minkowski never points out the link between his group-approach to a 4-dimensional geometry and Klein's famous Erlangen programme (which consisted in {\it defining} a geometry by its symmetry group, rather than by the `objects' on which it acts). This is all the more surprising since Klein was the organizer of the mathematics section in which Minkowski was invited to speak. Knowing also all what Minkowski owed to Felix Klein, I would have expected Minkowski to add at least a passing allusion to his Erlangen Programme. For instance, Pauli's famous article (and book) on Relativity contains a section (\S8) on how Relativity fits within Klein's ``Erlangen Programme'' \cite{Pauli21}. Maybe one should not, however, read too much into too little.

\subsection{No spacetime triangular inequalities}

I also find somewhat surprising the fact that Minkowski did not mention the issue of `triangular inequalities' in the 4-dimensional geometry. Indeed, Minkowski was a world expert on the generalizations of the usual triangular inequality based on convexity properties. The $L_p$-norm generalization of the usual ($L_2$-norm) triangular inequality, viz
\begin{equation}
\forall \ x,y \in {\mathbb C}^N \, , \ \Vert x+y \Vert_p \leq \Vert x \Vert_p + \Vert y \Vert_p \, , \ 1 \leq p \leq \infty
\end{equation}
is called ``the Minkowski inequality'' because of his fundamental contributions. Minkowski was also famous for his use of convexity inequalities in various fields of mathematics (geometry, number theory). The problem might be here that, as I mentioned above, Minkowski had (seemingly) not fully grasped the striking result of Einstein that the proper time along any polygonal (or curved) time-like line between two points in spacetime is {\it smaller} than the proper time along the straight line joining the two points. If he had realized it clearly, he would have commented that this is just the opposite of the usual triangular inequality, namely
\begin{equation}
\Vert x + y \Vert \geq \Vert x \Vert + \Vert y \Vert
\end{equation}
holds for future-directed, time-like vectors $x$ and $y$.

\section{Conclusion: Viva la Revolu\c cion!}
\setcounter{equation}{0}

To conclude these somewhat disconnected remarks, let me try to characterize the greatness of the conceptual leap achieved by Minkowski in his Raum und Zeit lecture by contrasting it with the attitude of Poincar\'e. We recalled above that, at the purely technical level, several (though certainly not all) of the key mathematical structures of ``Minkowski spacetime'' were already, explicitly or implicitly, contained in Poincar\'e's Rendiconti paper. But, what made the difference was that Minkowski had the boldness of realizing and publicizing the {\it revolutionary} aspects of these structures. The draft manuscripts of his lecture (see \cite{Galison}) show that Minkowski struggled in finding an appropriate way of conveying to a large audience his enthusiasm and his feeling of the revolutionary nature of the spacetime concept. In particular, as pointed out in \cite{Galison}, one of the first versions of the second sentence (``They [the new views of space and time] are radical.''), reads

\begin{quotation}
``Their character is mightily revolutionary, to such an extent that when they are completely accepted, as I expect they will be, it will be disdained to still speak about the ways in which we have tried to understand space and time.''
\end{quotation}

By contrast with this way of assuming and welcoming a revolutionary way of thinking, it is striking to read the introduction of the Rendiconti paper of Poincar\'e. There, in a rather roundabout way, he compares the efforts of Lorentz (and of himself) in trying to understand (without putting in doubt the usual ideas about space and time) the key role of the velocity of light $c$ (which would, ``if one would admit the postulate of relativity'', enter {\it both} electromagnetism and gravitation) to the efforts of Ptolemy, who tried to ``save phenomena'' without worrying about common epicyclic frequencies in the planets. He is getting near the (einsteinian) idea that one might need to change one's kinematical ideas about space and time measurements, but says:

\begin{quotation}
``Maybe it would suffice to give up this definition [of the equality of two lengths], to turn upside down the theory of Lorentz as completely as the Ptolemaic system has been turned upside down by Copernicus. If this happens one day, this will not mean that the efforts of Lorentz had been vain; because, whatever one might think, Ptolemy has not been useless to Copernicus. [car Ptol\'em\'ee, quoi qu'on en pense, n'a pas \'et\'e inutile \`a Copernic.]''
\end{quotation}

This citation clearly shows the deeply conservative bend of Poincar\'e in physics. He is happy to contribute to the Lorentz-Ptolemy programme, and he steps back from any move that might shake its kinematical foundations. Minkowski, by contrast, had a lot of ambition and self-confidence (not to say chutzpah), and was keen on breaking new ground in mathematical physics. Without fully understanding what Einstein had done, nor (at least initially) what Poincar\'e had already achieved, he was lucky to unearth elegant and deep mathematical structures that were implicitly contained in their (and others') work, and had the boldness to embrace with enthusiasm their revolutionary character. One must certainly admire him for this achievement, though one might regret his unfairness towards Poincar\'e. However, I hope to have convinced the reader that ``whatever one might think, Poincar\'e-Ptolemy has not been useless to Minkowski-Copernicus''.

\vglue 1cm

\noindent {\bf Acknowledgments.} I am grateful to John Stachel for informative (email) discussions about Einstein's thoughts on ``illusory time'', to Scott Walter for several relevant
 comments (and for the citation of Max Born),
and to Friedrich Hehl for inviting me to think (and write) about Minkowski's Raum und Zeit lecture, and for some useful
suggestions. I thank Ingrid Peeters for help in understanding and translating several German texts of (or about) Minkowski, David Blair and Jean Terri\`ere for useful advice about English style, and C\'ecile Cheikhchoukh for her care in latexing the manuscript.

\end{document}